\documentclass{ws-ijmpa}
\usepackage{graphicx}
\usepackage{amsmath}
\usepackage[latin1]{inputenc}
\usepackage{amsbsy}
\usepackage{color}
\usepackage{epsfig}
\usepackage{graphicx}
\usepackage{amsmath}
\usepackage{amssymb}
\usepackage{bbm}
\usepackage{amsbsy}
\usepackage{slashed}
\DeclareMathOperator{\tr}{tr}

\usepackage{xcolor}

\newcommand{\slsh}[1]{{\not \! #1}}

\usepackage[normalem]{ulem}  %\sout{}

\begin{document}
\title{Thermo-magnetic nonlocal NJL model in the real and imaginary time formalisms.}

\author{F. M\'{a}rquez}
\address{Santiago College, Avenida Camino Los Trapenses 4007, 
Lo Barnechea, Santiago, Chile. }
\author{R. Zamora}
\address{Instituto de Ciencias B\'asicas, Universidad Diego Portales, Casilla 298-V, Santiago, Chile.}
\address{Centro de Investigaci\'on y Desarrollo en Ciencias Aeroespaciales (CIDCA), Fuerza A\'erea de Chile, Santiago, Chile.}

\maketitle

\begin{abstract}
In this article we study a nonlocal Nambu--Jona-Lasinio (nNJL) model with a Gaussian regulator in presence of a uniform magnetic field. We take a mixed approach to the incorporation of temperature in the model, and consider aspects of both real and imaginary time formalisms. We include confinement in the model through the quasiparticle interpretation of the poles of the propagator. By working in the real time formalism and computing the spectral density function, we find that the effect of the magnetic field on the poles of the propagator can be entirely absorbed within the mean field value of the scalar field. The analytic structure of our propagator is then preserved in the weak magnetic field limit. The effect of the magnetic field in the deconfinement phase transition is then studied. It is found that, like with chiral symmetry restoration, magnetic catalysis occurs for the deconfinement phase transition. It is also found that the magnetic field enhances the thermodynamical instability of the system. We work in the weak field limit, i.e. $(eB)<5m_\pi^2$. At this level there is no splitting of the critical temperatures for chiral and deconfinement phase transitions.
\end{abstract}

\maketitle

\section{Introduction}

In recent years there has been a growing interest in studying the phase diagram of Quantum Chromodynamics (QCD) in the presence of a magnetic field. Particularly, the effect of the magnetic field in the critical temperature for chiral phase transition has been studied in lattice QCD \cite{Braguta,Delia,Bali1,Bali2,Bali3} and through several effective models for nonperturbative QCD \cite{Boomsma01,Boomsma01,Loewe1,Agasian,Fraga,Fraga2,Andersen,Nosotros01,renato1,renato2,renato3,renato4,renato5,Gamayun}. In effective models, temperature can be incorporated through either the real time formalism (RTF) or the imaginary time formalism (ITF). Most articles do so in the latter, since calculations are usually simpler. However, the physical meaning of the thermal propagator in ITF is not as straight forward as in the RTF. This makes for the physical interpretation of some results a nontrivial matter. In this article we have chosen a mixed approach that will allow us to take advantage of the simplicity of the ITF, as well as to make some physical remarks based on the RTF thermal propagator. We will use the ITF to determine the temperature evolution of the model and then use this data to study the temperature evolution of the RTF propagator.\\

One of the main limitations that effective models have when dealing with the QCD phase diagram is the absence of a confining model and, therefore, a deconfinement phase transition. To tackle this, it has been suggested that the Polyakov loop and the dressed Polyakov loop may act as order parameters for the deconfinement phase transition. This approach is frequently used in SDEs studies \cite{CSF:09,CSFM:09,CSFA:10} and in Nambu--Jona-Lasinio (NJL) models studies~\cite{KKH:09,TKM:10}. However, the dressed Polyakov loop is related to chiral symmetry restoration and, since both chiral and deconfinement phase transitions occur at similar temperatures \cite{EBF:08}, the validity of this approach in NJL models has been questioned \cite{Fede01}.\\

Nonlocal Nambu--Jona-Lasinio (nNJL) models offer an alternative approach to the realization of quark confinement. Through the incorporation of a nonlocal interaction, quark fields acquire a momentum-dependent dynamical mass. This means that the quark propagator may exhibit complex singularities with nonvanishing imaginary parts. Through the identification of the poles of the propagator as quasiparticles, one may interpret said quasiparticles to be confined when they exhibit a pole with a nonvanishing imaginary part \cite{Birse01,Birse02,Fede02}. This allows for a direct determination of the deconfinement critical temperature, as well as the determination of the dependence that mass and decay width have with temperature.\\

In order to be able to include confinement in the described manner in nNJL models one must work in Minkowski space, and therefore, in the RTF. It is in aspects like this that the value of RTF lies and our motivation for a mixed approach is based on. By computing the thermal propagator of a nNJL model in RTF, we can study the behavior of confined quasiparticles with temperature, and study the critical temperature for the deconfinement phase transition. On the other hand, the ITF is used to compute the thermal evolution of the order parameters, since it is simpler to do so in this formalism.\\

In a nonmagnetic nNJL model, deconfinement and chiral phase transition occur at a similar temperature. However, there are claims suggesting that, when in pressence of a uniform magnetic field, such transitions should decouple \cite{Fraga,Fraga2}. Our mixed aproach will therefore allow us to also comment on the splitting of chiral and deconfinement phase transitions.\\

The paper is organized as follows. In Sec. II, the Thermal nNJL model is introduced. In Sec. III we incorporate a uniform magnetic field and compute the real time thermal propagator. In Sec. IV the results of the investigation are presented. In Sec. V we present our conclusions.\\

\section{Thermal \lowercase{n}NJL Model.}

The nNJL model is described through the Euclidean Lagrangian
\begin{equation}\mathcal{L}_E=\left[\bar{\psi}(x)(-i\slashed{\partial}+m)\psi(x)-\frac{G}{2}j_a(x)j_a(x)\right],\label{lagrangiannjl}\end{equation}
with $\psi(x)$ being the quark field. The nonlocal aspects of the model are incorporated through the nonlocal currents $j_a(x)$
\begin{equation}j_a(x)=\int d^4y\,d^4z\,r(y-x)r(z-x)\bar{\psi}(x)\Gamma_a\psi(z),\end{equation} 
where $\Gamma_a=(1,i\gamma^5\vec{\tau})$ and $r(x)$ is the so-called regulator of the model. A bosonization procedure can be performed by defining scalar ($\sigma$) and pseudoscalar ($\vec{\pi}$) fields. Then, in the mean field approximation,
\begin{eqnarray}
\sigma&=&\bar{\sigma}+\delta\sigma\\
\vec{\pi}&=&\delta\vec{\pi},
\end{eqnarray}
where $\bar{\sigma}$ is the vacuum expectation value of the scalar field, serving as an order parameter for the chiral phase transition Also, it was assumed for the pseudoscalar field to have a null vacuum expectation value because of isospin symmetry. Quark fields can then be integrated out of the model \cite{Scoccola02,Scoccola04} and the mean field effective action can be obtained
\begin{equation}\Gamma^{MF}=V_4\left[\frac{\bar{\sigma}^2}{2G}-2N_c\int\frac{d^4q_E}{(2\pi)^4}\tr\ln S_E^{-1}(q_E)\right],\end{equation}
with $S_E(q_E)$ being the Euclidean effective propagator
\begin{equation}S_E=\frac{-\slashed{q}_E+\Sigma(q_E^2)}{q_E^2+\Sigma^2(q_E^2)}.\label{Euprop}\end{equation}
Here, $\Sigma(q_E^2)$ is the constituent quark mass
\begin{equation}\Sigma(q_E^2)=m+\bar{\sigma}r^2(q_E^2).\end{equation}
Finite temperature ($T$) effects can be incorporated through the ITF or Matsubara formalism. To do so, one can make the following substitutions
\begin{eqnarray}
V_4&\rightarrow&V/T\\
q_4&\rightarrow&-q_n\\
\int\frac{dq_4}{2\pi}&\rightarrow&T\sum_n,
\end{eqnarray}
where $q_n$ includes the Matsubara frequencies
\begin{equation}q_n\equiv(2n+1)\pi T.\end{equation}
With this, the propagator in Eq. (\ref{Euprop}) will now look like
\begin{equation}S_E(q_n,\boldsymbol{q},T)=\frac{\gamma^4q_n-\boldsymbol{\gamma}\cdot\boldsymbol{q}+\Sigma(q_n,\boldsymbol{q})}{q_n^2+\boldsymbol{q}^2+\Sigma^2(q_n,\boldsymbol{q})}.\label{Euprop2}\end{equation}
It is worth noting that the propagator in Eq. (\ref{Euprop2}) has no singularities. Since there are no poles at some $p^2$, the definition of an effective mass for the particle with such propagator is not clear and therefore the quasiparticle interpretation cannot be made.\\

The $\sigma$ field will evolve with temperature. This evolution can be computed through the grand canonical thermodynamical potential in the mean field approximation $\Omega_{MF}(\bar{\sigma},T,\mu)=(T/V)\Gamma_{MF}(\bar{\sigma},T,\mu)$ \cite{Kapusta01}. Then the value of $\bar{\sigma}$ must be at the minimum of the potential where 
$\partial\Omega_{MF}/\partial\bar{\sigma}=0$, which means
\begin{equation}
\left.\frac{\bar{\sigma}}{G}=2N_cT\sum_n\int\frac{d^3q}{(2\pi)^3}
r^2(q_E^2)\tr 
S_E(q_E)
\right|_{q_4=-q_n}.
\label{gap}
\end{equation}
Similar derivations of the gap equation in the ITF are readily available in the literature (see for example \cite{Scoccola02,Scoccola06}). We will use this equation to determine the temperature evolution of $\bar{\sigma}$. Since this will determine the temperature evolution of the model, $\bar{\sigma}(T)$ must behave in the same way regardless of the formalism used for the incorporation of temperature. This means that the behavior of $\bar{\sigma}(T)$ obtained in the ITF can be used within the RTF, since both formalisms must yield the same temperature behavior for the model. Therefore, we will also derive the real time thermal propagator in the model, that will then be studied using results obtained from the ITF.\\

To introduce RTF we must first perform a Wick rotation $q_4=iq_0$ that will take us from Euclidean to Minkowski space. Doing this in Eq. (\ref{Euprop}) will yield the zero temperature Minkowski space propagator
\begin{equation}S_0=i\frac{\slashed{q}+\Sigma(-q^2)}{q^2-\Sigma^2(-q^2)},\label{zp}\end{equation}
where $q^2=-q_E^2$. This propagator has singularities in the complex $q^2$ plane. Each of these singularities may be interpreted as a different quasiparticle and a mass and decay width may be defined. If $q^2=\mathcal{M}^2$ is a singularity of the propagator, the following definition can be made
\begin{equation}q^2=\mathcal{M}^2=M^2+iM\Gamma,\label{int}\end{equation} 
where $M$ is the constituent mass of the quasiparticle and $\Gamma$ its decay width \cite{Fede02,LoeweMorales}. We will therefore interpret complex poles with nonvanishing imaginary parts as confined quarks, and real poles as deconfined quarks. It is worth noting that a pole with a negative real part may occur in the nNJL model. These poles have been shown to produce thermodynamical instabilities and we will therefore avoid sets of parameters where this is the case \cite{Fede02}\\

In RTF, the number of degrees of freedom is doubled \cite{Ojima01,Ojima02,Kobes01,Landsman01,LeBellac01,Das01}. This means that the thermal propagator is given by a $2\times2$ matrix with elements $S_{ij}$. However, in one-loop calculations only the $S_{11}$ component is necessary. A general expression for $S_{11}$ can be written in terms of the spectral density function (SDF)
\begin{equation}S_{11}=\int\frac{dk_0}{2\pi i}\frac{\rho(k_0,\boldsymbol{q})}{k_0-q_0-i\varepsilon}-n_F(q_0)\rho(q),\label{prop}\end{equation}
where $n_F(q_0)$ is the Fermi-Dirac distribution $n_F(q_0)=({\rm{e}}^{q_0/T}+1)^{-1}$. The SDF can be obtained from \cite{LoeweMorales,Fede02,Fede03} 
\begin{equation}\rho(q)=S_+(q)-S_-(q), \label{rho}\end{equation}
where
\begin{equation}S_{\pm}(q)=\pm\oint_{\Gamma^{\pm}}\frac{dz}{2\pi i}\frac{S_0(z\mp i\varepsilon,\boldsymbol{q})}{z-q_0\pm i\varepsilon}.\label{S}\end{equation}
This is just a generalization of the free particle case where $\rho(q)=S_0(q_0+i\varepsilon,\boldsymbol{q})-S_0(q_0-i\varepsilon,\boldsymbol{q})$. The integration path $\Gamma^{\pm}$ is shown in Fig. 1.\\

\begin{figure}[!htb]
\begin{center}
\includegraphics[scale=0.4]{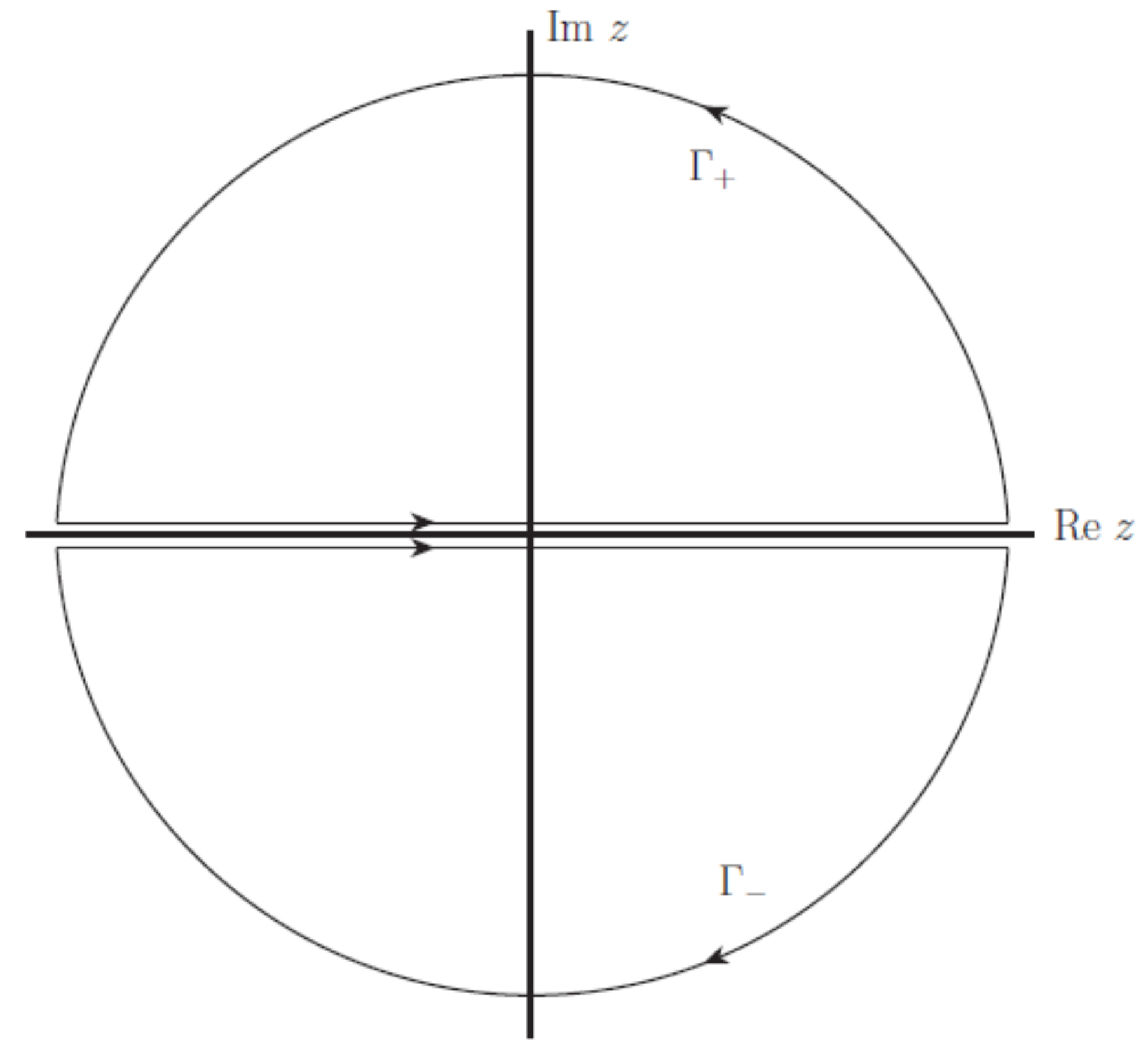}
\caption{Integration path in the definition of $S_{\pm}$.}
\label{pathS}
\end{center}
\end{figure}

The integrations can be performed and the SDF can be found to be 
\begin{equation}\rho(q)=\sum_{\mathcal{M}}i\left[\frac{A(\mathcal{M}^2)}{\mathcal{M}^2-q^2}-\frac{A((\mathcal{M}^2)^*)}{(\mathcal{M}^2)^*-q^2}\right]\label{SDF}\end{equation}
where the sum is over the various poles ($\mathcal{M}$) of the propagator and
\begin{multline}A(\mathcal{M}^2)=\frac{Z(\mathcal{M}^2)}{2E}\left(q_0(\slashed{q}+\Sigma(-\mathcal{M}^2))\right.\\\left.-\gamma^0(q^2-\mathcal{M}^2)\right),\label{Apolo}\end{multline}
with $E^2=\mathcal{M}^2+\boldsymbol{q}^2$ and where
\begin{equation}Z(\mathcal{M}^2)=\left.\left[\frac{\partial}{\partial q^2}\left(q^2-\Sigma^2(-q^2)\right)\right]^{-1}\right|_{q^2=\mathcal{M}^2},\label{Z}\end{equation}
is the renormalization constant. This calculation is fairly general and is valid for any regulator with real or complex poles. The RTF thermal propagator can then be obtained by putting Eq. (\ref{SDF}) into Eq. (\ref{prop}). The propagator will have the zero temperature contribution decoupled from the finite temperature one, i.e.
\begin{equation}S_{11}(q,T,\mu)=S_0(q)+\tilde{S}(q,T).\label{dec}\end{equation}
Of course, in the case where $\Gamma\rightarrow0$, i.e. when considering real poles, then the propagator reduces to the usual Dolan-Jackiw propagator \cite{Dolan01}
\begin{multline}S_{DJ}(q, M)=(\slashed{q}+M)\left[\frac{i}{q^2-M^2+i\varepsilon}\right.\\\left.-2\pi N(q_0)\delta(q^2-M^2)\right]\end{multline}
In this article we will use the gap equation in ITM to obtain the behavior of $\bar{\sigma}$ with temperature and then the RTF thermal propagator to gain physical insight from our results.\\

In the next section we will expand Eqs. (\ref{gap}) and (\ref{SDF}) to include a uniform magnetic field.\\

\section{Thermo-magnetic propagator in the real time formalism.}

We are interested in studying the model coupled to a homogeneous magnetic field. The derivative in the Lagrangian (\ref{lagrangiannjl}) is replaced by a covariant derivative
\begin{equation}
D_\mu=\partial_\mu + ie_fA_\mu.
\end{equation}
where $A^{\mu}$ is the vector potential corresponding to an homogeneous external magnetic field $\boldsymbol{B}=|\boldsymbol{B}|\hat{z}$ and $e_f$ is the electric charge of the quark fields (i.e. $e_u = 2e/3$ and $e_d = -e/3$). In the symmetric gauge,
\begin{equation}
A^{\mu}= \frac{B}{2}(0,-y,x,0),
\end{equation}
The Schwinger proper time representation for the propagator is given by \cite{Schwinger}
\begin{multline}
S(q)=-i\int_0^\infty ds \frac{e^{-is(M^2-q_{\|}^2+q_\perp^2
   \frac{\tan (eBs)}{eBs})}}{\cos (eBs)}
   \\\times\biggl[\left(\cos (eBs) + \gamma_1 \gamma_2 \sin (eBs)\right)
   (M+\slsh{q_{\|}}) - \frac{\slsh{q_\bot}}{\cos(eBs)} \biggr],
\end{multline} 
with $q_\|^2=q_0^2+q_3^2$, $q_\bot^2=q_1^2+q_2^2$ and where $e$ is the charge of the particle being $B$ the magnetic field.\\ 

For simplicity, we will consider the weak magnetic field case. The fermionic propagator in this region can be written as \cite{Taiwan}
\begin{eqnarray}
&&S(q) = i\frac{(\slsh{q}+\Sigma(-q^2))}{q^2-\Sigma^2(-q^2)} - \frac{\gamma_1 \gamma_2(eB) (\slsh{q}_{\|})+\Sigma(-q^2)}{(q^2-\Sigma^2(-q^2))^2} \nonumber \\
&-&\frac{2 i(eB)^2 q_{\bot}^2}{(q^2-\Sigma^2(-q^2))^4} \nonumber \\ 
&\times&\biggl[ (\Sigma(-q^2)+\slsh{q}_{\|}) 
+ \frac{\slsh{q}_{\bot}(\Sigma^2(-q^2)-{q}_{\|}^2)}{q_{\bot}^2} \biggr]. \label{debil}
\end{eqnarray}

We will now turn to the calculation of the spectral density function in presence of a magnetic field. To do this, we have to compute Eqs. (\ref{rho}) and (\ref{S}) using the propagator in Eq. (\ref{debil}). The integration can be performed through residues to obtain

\begin{eqnarray}\rho(q)&=&\sum_{\mathcal{M}}2i\left[\frac{A(\mathcal{M}^2)}{\mathcal{M}^2-q^2}-\frac{A((\mathcal{M}^2)^*)}{(\mathcal{M}^2)^*-q^2}\right] \nonumber \\
&+& (eB)\gamma_1\gamma_2\left[\frac{B(\mathcal{M}^2)}{(\mathcal{M}^2-q^2)^2}-\frac{B((\mathcal{M}^2)^*)}{((\mathcal{M}^2)^*-q^2)^2}\right] \nonumber \\
&-& 2i(eB)^2 \left[\frac{C(\mathcal{M}^2)}{(\mathcal{M}^2-q^2)^4}-\frac{C((\mathcal{M}^2)^*)}{((\mathcal{M}^2)^*-q^2)^4}\right], \nonumber \\
\label{SDFcampo}\end{eqnarray}
where the sum is over the various poles ($\mathcal{M}$) of the propagator, $A(\mathcal{M})$ and $Z(\mathcal{M})$ are defined in Eqs. (\ref{Apolo}) and (\ref{Z}) respectively. Also 
\begin{equation}
B(\mathcal{M}^2) = Z^2(\mathcal{M}^2) (\slashed{q}_{\|}+\Sigma(-\mathcal{M}^2) ),
\end{equation}
and
\begin{eqnarray}
&&C(\mathcal{M}^2) =\frac{Z^4(\mathcal{M}^2)}{E^7}(5 E^8(-\gamma_0 q_{\bot}+q_0 \slashed{q}_{\bot})) \nonumber \\
&+& 5 q_0^7 (-\slashed{q}_3 q_{\bot}^2
+q_{3}^2 \slashed{q}_{\bot} + \Sigma(-\mathcal{M}^2) q_{\bot}^2  + \Sigma^2(-\mathcal{M}^2) \slashed{q}_{\bot} ) \nonumber \\
&-&5 E^6q_0 (3 \slashed{q}_{0} q_{\bot}^2- \slashed{q}_{3} q_{\bot}^2 -3q_{0}^2 \slashed{q}_{\bot} + 3q_{3}^2 \slashed{q}_{\bot}
 \nonumber \\ &+&7 \Sigma(-\mathcal{M}^2) q_{\bot}^2 + \Sigma^2(-\mathcal{M}^2) \slashed{q}_{\bot} )  \nonumber \\
 &+& E^2 q_{0}^5 (-\slashed{q}_{0} q_{\bot}^2+21 \slashed{q}_{3} q_{\bot}^2-21\Sigma(-\mathcal{M}^2) q_{\bot}^2 
 +\slashed{q}_{\bot} q_{0}^2 \nonumber\\
 &-&21\slashed{q}_{\bot} q_{3}^2 -21\Sigma^2(-\mathcal{M}^2) \slashed{q}_{\bot}) \nonumber \\
 &+& 5 E^4q_{0}^3 (\slashed{q}_{0} q_{\bot}^2-7 \slashed{q}_{3} q_{\bot}^2+7\Sigma(-\mathcal{M}^2) q_{\bot}^2 
 -\slashed{q}_{\bot} q_{0}^2 \nonumber\\
 &+&7\slashed{q}_{\bot} q_{3}^2 +7\Sigma^2(-\mathcal{M}^2) \slashed{q}_{\bot}).
\end{eqnarray}

From this we can see that the singularities of the propagator will still be found at $q^2=\mathcal{M}^2$. All of the contributions of the magnetic field then has been absorbed through the dependence of $\bar{\sigma}$ on the magnetic field and therefore the interpretation made in Eq. (\ref{int}) holds true for the nonvanishing magnetic field case.

\section{Results.}

Throughout this work, we will use the Gaussian regulator for the nNJL model, i.e.
\begin{equation}r^2(-q^2)={\rm e}^{-q^2/\Lambda^2}.\end{equation}
For the parameters of the model, we take \cite{LoeweMorales} $m=10.5$ MeV, $\Lambda=627$ MeV and $G=5\times10^{-5}$ MeV$^2$. With this set of parameters we have $\bar{\sigma}_0=339$ MeV.\\

\begin{figure}[!htb]
\begin{center}
\includegraphics[scale=0.4]{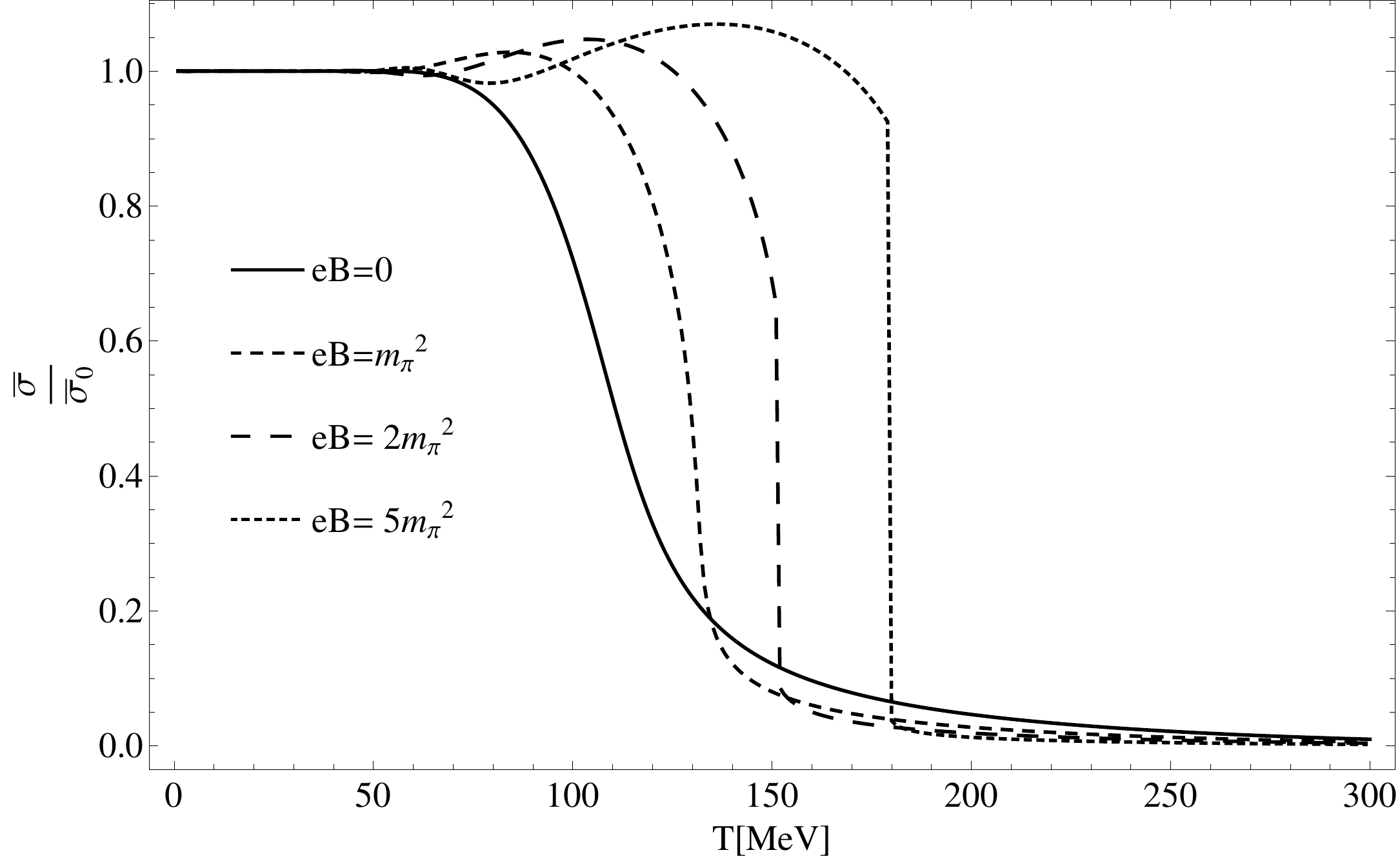}
\caption{Temperature dependence of $\bar{\sigma}$ for different magnetic fields. $(\bar{\sigma}_0)^*$ is the mean field value of the scalar field at $T=0$ and $B=0$.}
\label{SigmaB}
\end{center}
\end{figure}

Figure \ref{SigmaB} shows the behavior of $\bar{\sigma}$ with temperature for different values of the magnetic field. As expected, at $(eB)=0$, rather than a phase transition, we have a {\em cross-over} from the chirally broken to the chirally restored phase. However, as the magnetic field increases the shape of the transition becomes similar to a first order phase transition, in accordance to what was found in \cite{Kashiwa}.\\

Also, as the magnetic field increases, Fig. \ref{SigmaB} shows a growth of $\bar{\sigma}$ with temperatures below the critical temperature. This sort of behavior has been proven to relate to negative pressure and oscillating entropy in nNJL models \cite{Blaschke04,Blaschke05}. In this manner, the magnetic field has an undesirable effect on the thermodynamical behavior of the model. This can be further studied by looking at the poles of the propagator.\\

\begin{figure}[!htb]
\begin{center}
\includegraphics[scale=0.45]{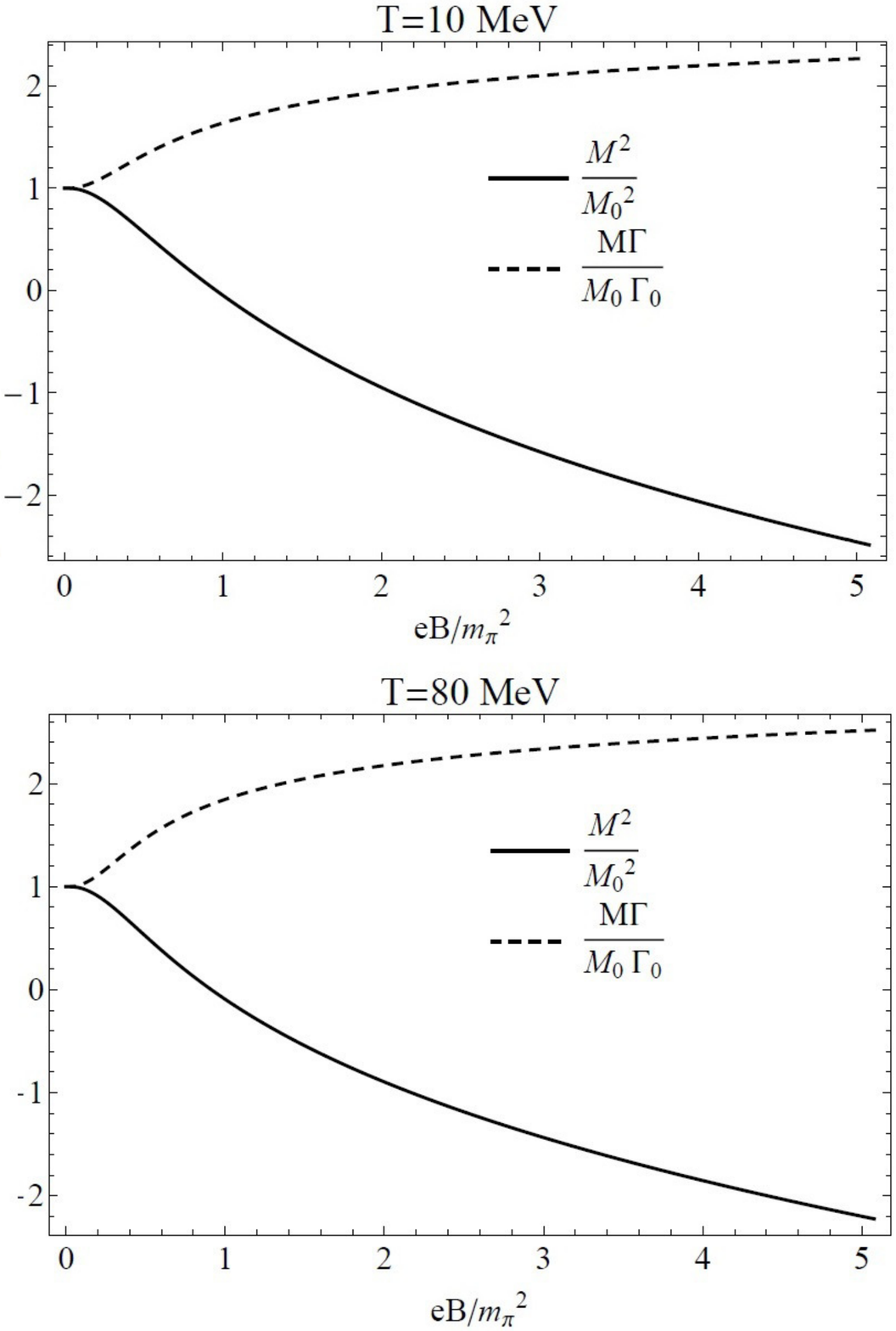}
\caption{Real and imaginary parts of the first pole of the propagator, as a function of $(eB)$ for different temperatures. The plot on the left is at $T=10$ MeV and the plot on the right at $T=80$ MeV.  $M_0$ and $\Gamma_0$ $M_0$ and $\Gamma_0$ are the mass and decay width at $B=0$, according to Eq. (\ref{int}).}
\label{MGamma}
\end{center}
\end{figure}

Figure \ref{MGamma} shows the behavior of the real and imaginary parts of the first pole of the propagator as a function of temperature. Since the first pole is also the lighter one, the behavior of the system is dominated by this pole and the contribution from other poles can be (somewhat) neglected. The imaginary part grows with the magnetic field and the real part decreases. If we follow the quasiparticle interpretation made in Eq. (\ref{int}), this means the quasiparticle is highly unstable. This kind of poles have been shown to produce rising values for the chiral condensate and $\bar{\sigma}$ \cite{Fede02} as a function of temperature, which in turn accounts for negative pressure and oscillating entropy. The magnetic field then, makes our system more thermodynamically unstable.\\

This same quasiparticle interpretation allows us to define both a confined and a deconfined phase and therefore a critical temperature for the deconfinement phase transition. When the imaginary part of the pole is finite we will say we are in the confined phase. When the imaginary part of the pole vanishes we will say we are in the deconfined phase. The temperature which divides one regime from the other is then the critical temperature for the deconfinement phase transition.\\

\begin{figure}[!htb]
\begin{center}
\includegraphics[scale=0.65]{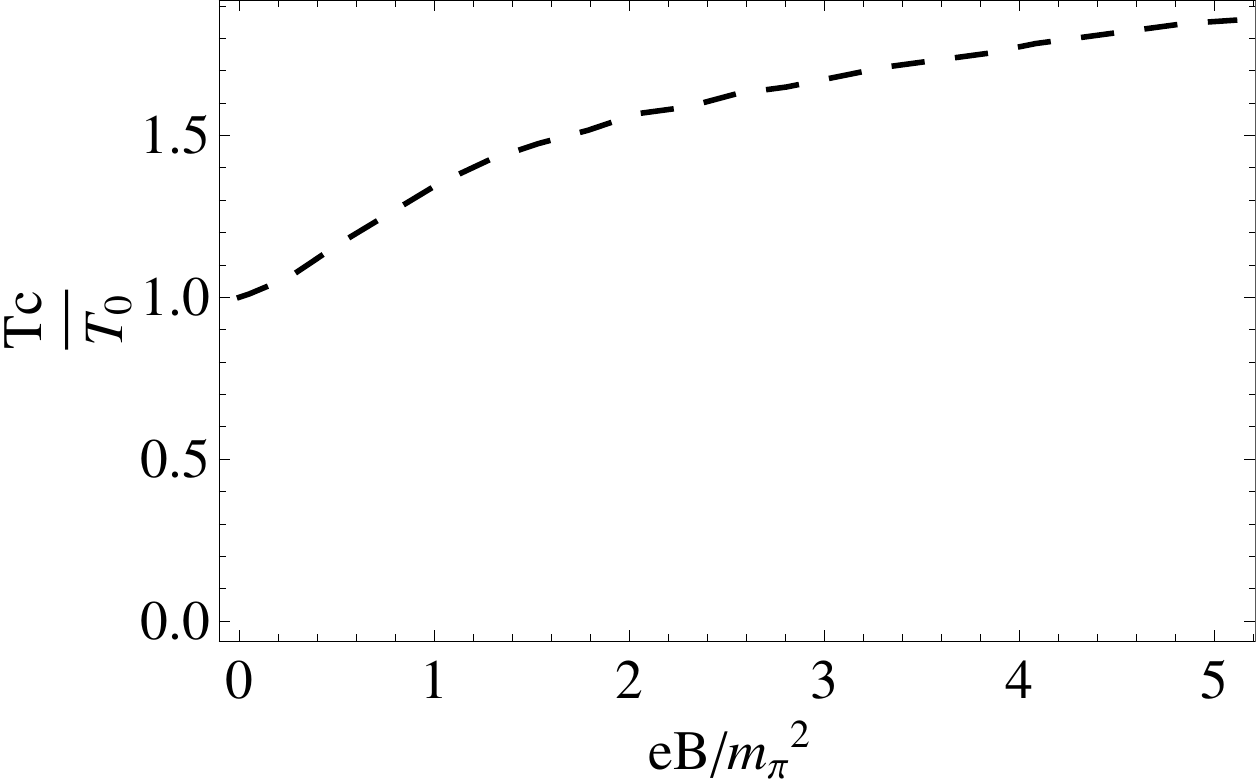}
\caption{Critical temperature for the deconfinement phase transition as a function of the magnetic field. $T_0$ is the deconfinement critical temperature at $B=0$.}
\label{Tcrit}
\end{center}
\end{figure}

Figure \ref{Tcrit} shows the behavior of the critical temperature for the deconfinement phase transition as a function of the magnetic field. There is a clear magnetic catalysis for the deconfinement critical temperature. From Fig. \ref{SigmaB} we can see that there is also a magnetic catalysis for the chiral phase transition critical temperature.\\

\begin{figure}[!htb]
\begin{center}
\includegraphics[scale=0.6]{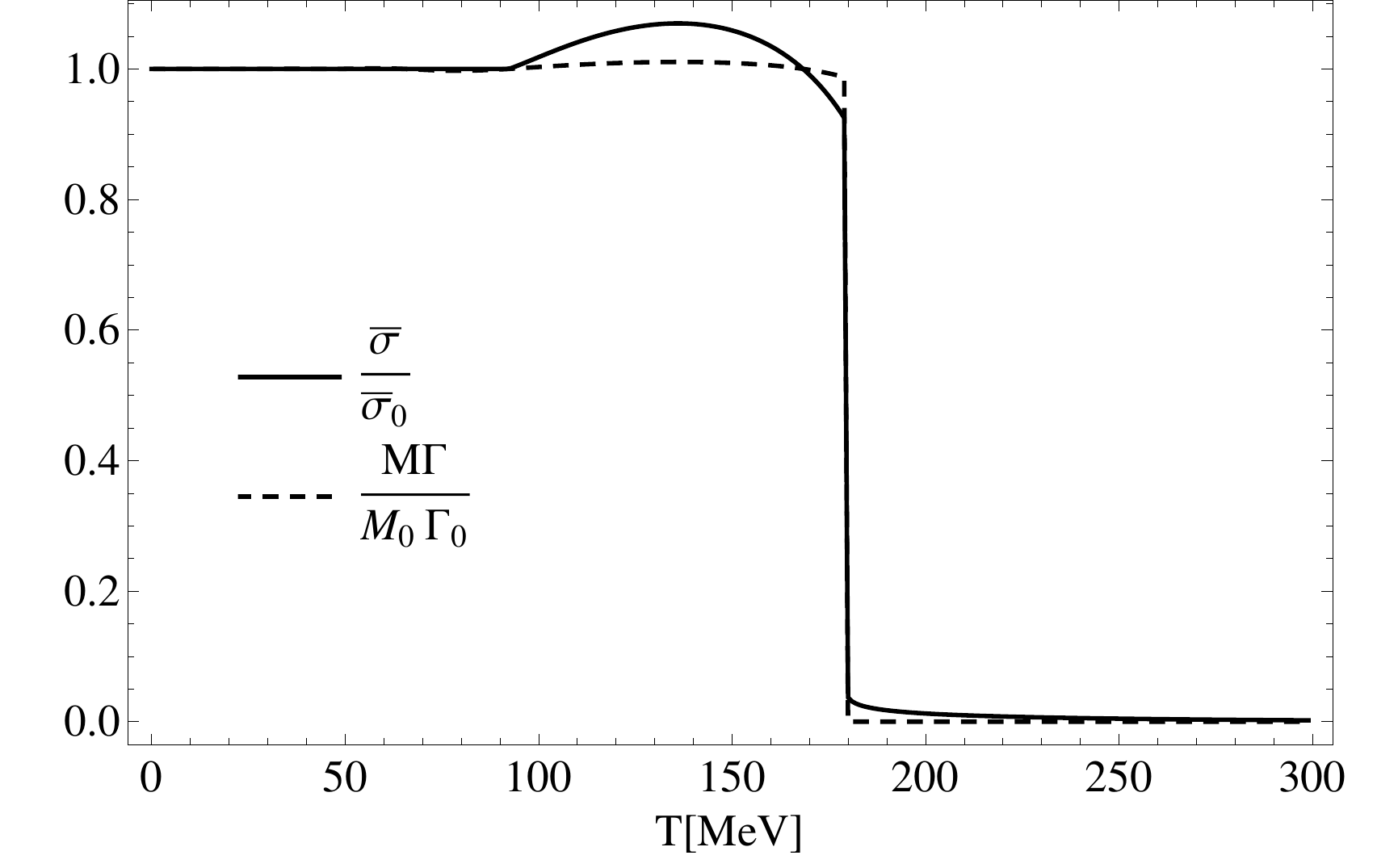}
\caption{$\bar{\sigma}$ and the imaginary part of the pole as a function of temperature for $(eB)=5m_\pi^2$. $\bar{\sigma}_0$ is the mean field value of the scalar field at $T=0$. $M_0$ and $\Gamma_0$ are the mass and decay width at $B=0$, accordin to Eq. (\ref{int})}
\label{Split}
\end{center}
\end{figure}

Figure \ref{Split} shows the behavior of both $\bar{\sigma}$ and  the imaginary part of the pole with temperature. Both quantities fall to zero at the same temperature, signaling a first order phase transition for both confinement and chiral symmetry. This means that, at least at this level, there is no splitting of the chiral and deconfinement phase transitions.\\

\section{Conclusions.}

In this article we studied the thermodynamical behavior of a nNJL model in presence of a weak magnetic field. The real time thermal propagator in the weak field limit was computed and it exhibits the same poles as the thermal propagator in absence of a magnetic field. This allows to study the thermodynamical stability of the system, through the quasiparticle interpretation, in presence of a magnetic field.\\

The imaginary part of the poles of the propagator grows with the magnetic field, making the quasiparticles more unstable and therefore trigering the thermodynamical instability of the system.\\

As the magnetic field increases, the chiral phase transition turns form a cross-over to a first order phase transition. Also, as the field increases, the thermodynamical behavior of the model is compromised as seen by the growth of $\bar{\sigma}(T)$ before the chiral phase transition. This sort of behavior is caused by unstable quasiparticles, also triggered by the magnetic field.\\

Finally we studied the behavior of the critical temperature for the deconfinement phase transition with the magnetic field. We found that magnetic catalysis takes place and, furthermore, there is no splitting of the transitions for the considered values of the magnetic field. However, there is no claim to be made as to what may be found at higher values of the magnetic field. \\

It is worth noting that this analysis has been carried out in the mean field approximation, therefore, it may be the case that going beyond mean field different phenomenology may be found. It has been shown, for example, that a magnetic field dependent coupling constant in effective models may yield an inverse magnetic catalysis in the chiral phase transition \cite{renato4,renato5}. Re-summation effects, next-to-leading order calculations and consider effect of gauge interactions may also have a nontrivial effect on both chiral phase transition \cite{renato4,renato5,Contrera:2009hk,Pagura:2016pwr,Pagura:2016rit,Dumm:2013zoa,Pagura:2012ku}. Going beyond mean field, one should consider several different aspects as new interactions, corrected couplings and re-summation of Feynman diagrams \cite{Buballa01,Buballa02,Buballa03,Buballa04}. The incorporation of such effects, however goes beyond the scope of the present article and is left as future work to be considered.

\section{Acknowledgements}

R. Zamora would like to thank support from CONICYT FONDECYT Iniciación under grant No. 11160234.

\bibliography{BNJL}{}
\bibliographystyle{ws-ijmpa}

\end{document}